\documentclass{moriond}

\usepackage{xspace}
\usepackage{siunitx}

\def\be{\begin{equation}}
\def\ee{\end{equation}}
\def\bea{\begin{eqnarray}}
\def\eea{\end{eqnarray}}

\newcommand{\Photo}{\includegraphics[height=35mm]{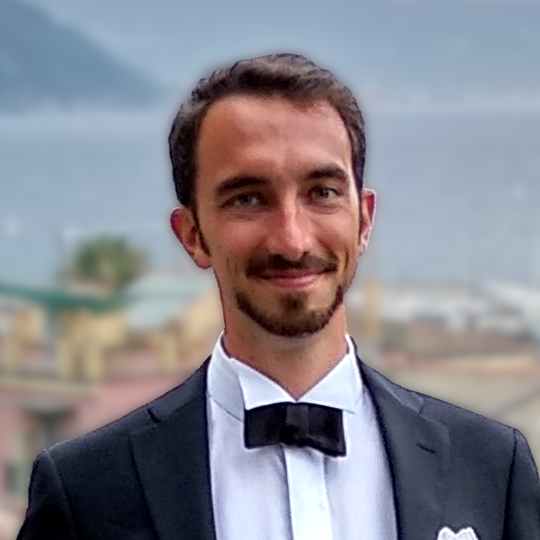}}

\begin{document}

\newcommand{\aegis}{AE$\overline{\textrm{g}}$IS\xspace}
\newcommand{\Hbar}{$\overline{\textrm{H}}$\xspace}
\newcommand{\pbar}{$\overline{\textrm{p}}$\xspace}
\newcommand{\pos}{$\mathrm{e^+}$\xspace}
\newcommand{\ele}{$\mathrm{e^-}$\xspace}


\vspace*{4cm}
\title{Progress towards measuring the fall of antimatter in Earth’s gravitational field}

\author{Marco Volponi, on behalf of the AEgIS Collaboration}

\address{CERN, Esplanade des Particules 1, Geneva, Switzerland \\ Dipartimento di Fisica, Università di Trento, via Sommarive 14, 38123 Povo (TN), Italy}

\maketitle\abstracts{
The \aegis (Antimatter Experiment: Gravity, Interferometry, Spectroscopy) experiment, located at the Antimatter Factory at CERN, aims to study the asymmetry between matter and antimatter. In particular, its first goal is to measure the effect of gravity on antimatter. The method chosen is to determine the fall of a pulsed beam of anti-hydrogen, caused by the Earth’s gravitational field, by mean of a moiré deflectometer. In this contribution the new anti-hydrogen production scheme is presented, together with the improvements that the experimental setup underwent in the last years, deemed necessary in order to reach an anti-hydrogen flux with the characteristics needed to obtain a precise gravity measurement. Last, the first technical results are described and the future steps outlined.
}

\section{Introduction}

The discrepancy between the abundance of matter over antimatter in the observable universe has long puzzled physicists, as the Standard Model predicts that equivalent quantities of both should have been generated during the Big Bang. A difference in the coupling of matter and antimatter with gravity would violate the Weak Equivalence Principle, and could be an explanation for this asymmetry. Certain theories beyond the Standard Model admit such a discrepancy~\cite{nieto_goldman}. 
The principal objective of the \aegis (Antimatter Experiment: gravity, Interferometry, Spectroscopy) experiment is to measure the acceleration of antimatter in Earth's gravitational field. The technique is based on producing a precisely timed, pulsed beam of anti-hydrogen atoms (\Hbar), allowing it to free fall through the gratings of a moiré deflectometer, and observing the interference pattern on the detector to infer the acceleration due to gravity.

\section{\aegis\ Phase 1: anti-hydrogen pulsed production demonstration}
\label{sec:aegis1}

The main target of \aegis Phase 1 was to produce, in a pulsed manner, \Hbar with antiprotons (\pbar) from the Antimatter Decelerator (AD) and Rydberg Positronium atoms (Ps), using a charge-exchange reaction. The \pbar from AD were  decelerated from \SI{5.3}{\MeV} to below \SI{10}{\keV} using a \SI{150}{\um} aluminium foil, so to be trappable in a Penning trap. Afterwards, they were cooled by a combination of sympathetic cooling with \ele and compressing the \pbar plasma using the rotating wall technique~\cite{pbar_compression}. Then, a cloud of Ps moving perpendicularly in the direction of the \pbar plasma was generated by impinging a beam of keV positrons (\pos) on a nano-channelled silica porous \pos-Ps converter. The Ps atoms were excited to $n_{Ps}$ = 17 thanks to a two-step laser process~\cite{ps_vel}.
The formation of \Hbar was established by observing the variation in annihilation rates on a scintillator when antiprotons, positrons, and lasers were employed concurrently or separately. The production rate was estimated to be 0.05 \Hbar per experiment cycle~\cite{hbar_prod}, which lasted approximately \SI{110}{\s}.

The adopted \Hbar production strategy offers many benefits. Firstly, the production process is pulsed and precisely defined in time, with an uncertainty of $\sim$ \SI{250}{\ns}, allowing accurate time-of-flight analyses. Secondly, the temperature of the produced \Hbar is determined by the \pbar one, leading to formation of cold \Hbar. Additionally, the reaction cross-section, following $\sigma \propto n^4_{Ps}$, increases rapidly with $n_{Ps}$. 

\section{\aegis\ Phase 2: more anti-hydrogen and first gravity test}
\label{sec:aegis2}

The primary objectives of \aegis Phase 2 are enhancing and solidifying the process of anti-hydrogen formation, while simultaneously conducting the initial proof-of-concept gravitational measurement using antimatter. \aegis Phase 2 will span until 2025, with the start of CERN LS3. The findings from Phase 1 highlighted the need for a substantial increase of 3-4 orders of magnitude in \Hbar flux, aiming to O(10) \Hbar per cycle, in order to acquire the necessary statistic for gravity measurement with a precision level of $\sim$ 1\%. Furthermore, the cooling of \pbar's (and consequently \Hbar's) to temperatures on the order of tenths of kelvin, which is approximately one order of magnitude lower than the previous $\sim$ \SI{400}{\kelvin}, is crucial. The two additional milestones for the Phase 2 are the creation of a forward-boosted anti-hydrogen beam and the design, build, and test of a prototype moiré Deflectometer specifically tailored for gravity measurements.

To achieve this improvement, multiple modifications have been implemented in the apparatus. The most significant change is an enhanced \Hbar formation scheme: the \pos-Ps target is now placed along the trap's axis, enabling the Ps atoms to move collinearly with the magnetic field towards the \pbar plasma~\cite{moriond2022}. Consequently, the maximum achievable value of $n_{Ps}$ goes from 19 to over 30, surpassing the previous limit dictated by Ps ionization from the motional Stark effect ($n_{Ps}^{max}\propto \theta^{-1/4}$)\footnote{With $\theta$ we indicate the angle between the trajectory of the particle and the magnetic field.}. To exploit this capability, the crystal of the Ps second excitation laser has been changed, so to arrive to $n_{Ps}$ = 24, giving a 4-fold increase in the \Hbar formation cross-section. 
To accommodate the new scheme, the formation trap had to be redesigned, and it was produced and installed in 2022. The new trap features larger, fully circular electrodes and incorporates two-stage electrical noise filtering to enhance the stability and lifetime of the \pbar plasma.

Furthermore, starting from 2021, the \pbar's for the experiment has been given by ELENA (Extra Low Energy Antiproton ring)~\cite{ELENA}, with an energy down from the previous \SI{5.3}{\MeV} to \SI{100}{\keV}; ELENA simultaneously caters up to 4 experiments, delivering \pbar's in packets of $\sim 5 \times 10^6$ every 2 minutes. To optimize the transmission of \pbar with energies below \SI{14}{\keV}, which is the maximum allowed by our catching trap, new degraders have been installed. This combination  is expected to yield a minimum of 5 times more trappable \pbar per cycle compared to the 2018 setup, resulting in a significantly higher production of \Hbar.

The implementation of ELENA has also resulted in a significant change in the operational approach. Previously, the experiment required constant monitoring by multiple experts during 8-hour data-taking shifts; however, the current continuous beam delivery from ELENA would poses an unbearable workload for the operators. As a result, a paradigm shift has occurred, transitioning from a software infrastructure consisting of separate subsystems on different machines to a new comprehensive control system framework called TALOS (Total Automation of LabVIEW™ Operations for Science).
TALOS integrates all individual control programs into a coordinated distributed environment, thereby improving reliability and safety via multiple distributed watchdog systems, which guarantee that any unresponsive component is promptly detected. These features are fundamental to enable the full automation of experimental procedures and extended periods of unmanned operation, especially since high-level choices often rely on parameters generated by different PCs. The framework is modular, being constituted by several independent components, each with a specific purpose, operating asynchronously and communicating with one another through a built-in messaging system.

Simultaneously, the previous custom-made control electronics have been transitioned to a new system based on ARTIQ/Sinara~\cite{sinara}, which is an open hardware and software ecosystem specifically designed for quantum physics experiments. This hardware offers modularity and precise nanosecond-level synchronization capabilities, while ARTIQ, based on Python, eases the programming. This combination has rationalised the experimental procedures and further enhanced their timing accuracy.

Furthermore, a detailed study was conducted to optimise the \pos-Ps silica target by finely adjusting the morphology of the nano-channels~\cite{nanochannel}. Additionally, the ability to bake the target within the trap under cryogenic conditions has been incorporated. These advancements are expected to increase the Ps conversion efficiency by a factor of 5.

\section{Results from the 2022 data taking}
\label{sec:results}

In 2022 the assembly of the upgrades was completed, and the \pbar beam time was used to test, step by step, the performances of the improved apparatus. First, the beam steering has been optimised, by running an automated scan over the 4 tunable parameters (horizontal and vertical offsets and angles), and analysing the resulting image acquired with an MCP at the end of the experiment: with the best settings, more than 80\% of the \pbar's were fitting inside the hole of the thin degrader foils. After that, the catching optimisation was performed, by scanning over the electrodes opening and closing times, and their voltages. The catching performance was then estimated in two ways: by measuring all the losses with multiple scintillators, and subtracting it from the incoming quantity given by the AD operators, and by an absolute \pbar counting, a destructive measurement performed dumping the trapped fraction onto a Faraday Cup. The analyses on both measurements has still to be finalised, but a preliminary estimation gives two numbers compatible with a trapped fraction of $\sim$ 70\% of the incoming \pbar's. \aegis, in 2022, has therefore achieved the most efficient \pbar trapping from ELENA, routinely catching $\sim$ 3.7e6~\pbar's per bunch.

\begin{figure}[h]
\centerline{\includegraphics[width=0.9\linewidth]{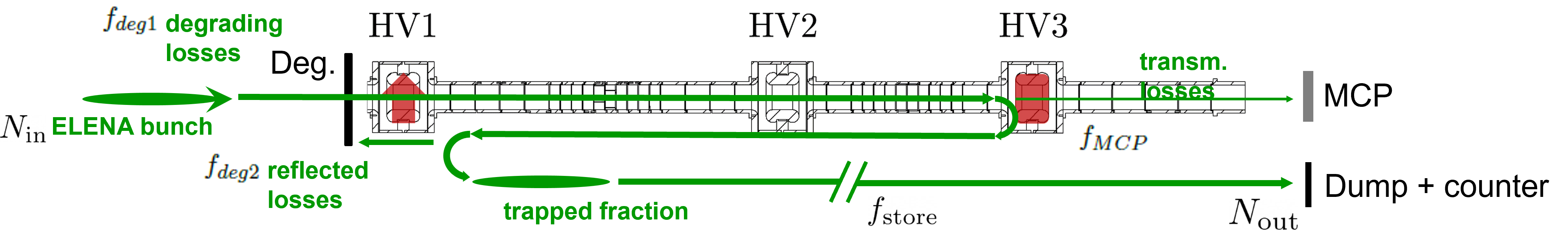}}
\caption[]{Schematics of the \pbar's capture, together with the parameters used to estimate their number.}
\label{fig:pbar_capture}
\end{figure}

\section{Conclusions and future development}

All these upgrades together will increase the \Hbar production of 3-4 orders of magnitudes, as shown in Table~\ref{tab:exp}. Some of them had to be postponed to 2024 due to delays in the supply chains, so during 2023 the expected \Hbar production rate will remain on the order of O(10)~\Hbar/min. In 2024 both $n_{Ps}$ will be raised above 30, and the \pos source will be renovated, so to further augment the formation to O(100)~\Hbar/min: this means that, integrated over an entire year of data taking, \aegis will produce millions \Hbar's.

The next step is to design a moiré deflectometer suitable for measuring the tiny deviation due to gravity. With a detector \SI{20}{\cm} long, with gratings' width of \SI{40}{\um}, a total of $\sim$ 1000 \Hbar's should be detected~\cite{gravimeter}. In the hypotheses of producing the \Hbar with $T \sim 10 K$, this amount of \Hbar detected would lead to a precision of \SI{\pm3}{\m\s^{-2}}, enough to constrain the sign of gravity on antimatter.

During LS3 (i.e. 2026$\div$2028) the \aegis apparatus will be further improved to both enhance formation rate and produce a colder anti-hydrogen source, to ultimately perform a gravity measurement with the precision of 1\%.

\begin{table}[t]
\caption[]{Anti-hydrogen formation improvement.}
\label{tab:exp}
\vspace{0.4cm}
\begin{center}
\begin{tabular}{llllll}
\hline
							& 2018 					& 2023 					& Gain 					& 2024 				& Gain 					\\
\hline
Degrader efficiency 		& 2-3\% $\times$ 3e7 	& 70\% $\times$ 5e6 	& 6 $\times$ 			& 					&						\\
\Hbar Form. Cross-section 	& N = 17 				& N = 24 				& 4 $\times$ 			& N $\geq$ 30			& 16 $\times$			\\
Positronium target 			& 7 \% 					& 25 \% 				& 3.6 $\times$			& 					&						\\
Laser coverage				& 15 \%					& 30 \%					& 2 $\times$			& 					&						\\
\pbar plasma density wrt. Ps& Perpendicular			& Collinear				& 2 $\times$			& 					&						\\
New \pos source				& 1e6~\pos/min			& 1e6~\pos/min			& 1 $\times$			& 8e6~\pos/min		& 8 $\times$			\\
\hline
\Hbar prod. gain			& 0.05~\Hbar/min			& O(10)~\Hbar/min		& $\sim$ 350 $\times$	& O(100)~\Hbar/min	& $\sim 10^4 \times$	\\
\hline
\end{tabular}
\end{center}
\end{table}

\section*{Acknowledgments}

This work has been performed in the framework of the \aegis\ Collaboration and it was supported by Istituto Nazionale di Fisica Nucleare (INFN-Italy), the CERN Fellowship and Doctoral Student program, and the University of Trento.

\end{document}